\documentclass[twocolumn]{aastex631}
\graphicspath{{./fig/}{./png/}}
\usepackage{bm}

\newcommand{\bec}[1]{\mbox{\boldmath $ #1$}}
\newcommand{\meanN}{\overline{n}}
\newcommand{\meanB}{\overline{B}}
\newcommand{\meanBB}{\overline{\bm B}}
\def\kappat{\kappa_{\rm t}}
\def\etat{\eta_{\rm t}}
\def\urms{u_{\rm rms}}
\def\kf{k_{\rm f}}
\def\Ma{\mbox{\rm Ma}}
\def\Rey{\mbox{\rm Re}}
\begin{document}

\title{Theory of the kinetic helicity effect on turbulent diffusion of magnetic and scalar fields}
\email{brandenb@nordita.org}
\author[0000-0001-7308-4768]{Igor Rogachevskii}
\affiliation{Department of Mechanical Engineering, Ben-Gurion University of the Negev, Beer-Sheva 84105, P. O. Box 653, Israel}
\affiliation{Nordita, KTH Royal Institute of Technology and Stockholm
University, Hannes Alfv\'ens v\"ag 12, SE-10691 Stockholm, Sweden}
\author[0000-0002-5744-1160]{Nathan Kleeorin}
\affiliation{Department of Mechanical Engineering, Ben-Gurion
University of the Negev, Beer-Sheva 84105, P. O. Box 653, Israel}
\affiliation{IZMIRAN, Troitsk, 108840 Moscow Region,  Russia}
\author[0000-0002-7304-021X]{Axel Brandenburg}
\affiliation{Nordita, KTH Royal Institute of Technology and Stockholm University, Hannes Alfv\'ens v\"ag 12, SE-10691 Stockholm, Sweden}
\affiliation{The Oskar Klein Centre, Department of Astronomy, Stockholm University, AlbaNova, SE-10691 Stockholm, Sweden}
\affiliation{McWilliams Center for Cosmology \& Department of Physics, Carnegie Mellon University, Pittsburgh, PA 15213, USA}
\affiliation{School of Natural Sciences and Medicine, Ilia State University, 3-5 Cholokashvili Avenue, 0194 Tbilisi, Georgia}

\begin{abstract}
Kinetic helicity is a fundamental characteristic of astrophysical turbulent flows.
It is not only responsible for the generation of large-scale magnetic fields in the Sun, stars, and spiral galaxies, but
it also affects turbulent diffusion, resulting in the dissipation of large-scale magnetic fields.
Using the path integral approach for random helical velocity fields with a finite correlation time and large Reynolds numbers,
we show that turbulent magnetic diffusion is reduced by the kinetic helicity, while
the turbulent diffusivity of a passive scalar is enhanced by the helicity.
The latter can explain the results of recent numerical simulations for forced helical turbulence.
One of the crucial reasons for the difference between the kinetic helicity effect on magnetic and scalar fields is related to the helicity dependence of the correlation time of a turbulent velocity field.
\end{abstract}

\section{Introduction}
\label{sec1}

The evolution of solar and Galactic large-scale magnetic fields can be understood in terms of mean-field dynamo theory, applying various analytical methods
\citep[see, e.g.,][]{M78,P79,KR80,ZRS83,RSS88,RHK13,MD19,RI21,SS21}.
Helical motions emerge in inhomogeneous or density-stratified turbulence, give rise to an $\alpha$ effect, and produce large-scale dynamo action
in combination with a nonuniform (differential) rotation,
while turbulent magnetic diffusion limits the growth rate of the field.

It has recently been shown using direct numerical simulations (DNS) \citep{BSR17,BKRY25}
that helical turbulent motions of the plasma affect not only the $\alpha$ effect, but also the turbulent magnetic diffusion.
In particular, the kinetic helicity $H_{\rm K} =\langle{\bm u} \cdot  {\bm \omega}  \rangle$
was found to lower the turbulent magnetic diffusion coefficient $\eta_{_{\rm t}}$,
where ${\bm u}$ and ${\bm \omega}$ are fluctuations of velocity and vorticity, and angular brackets denote ensemble averaging.
On the other hand, DNS showed that the kinetic helicity increases the turbulent diffusion coefficient for passive scalars \citep{BKRY25}.

Using the renormalization group approach in the limit of low magnetic Reynolds numbers,
it has been recently shown by \cite{MI23} that the decrease of the turbulent magnetic diffusion coefficient
in comparison with that for a nonhelical random flow
is of the order of ${\rm Rm}^2 (H_{\rm K} \tau_{\rm c})^2 / \langle {\bm u}^2 \rangle$,
where ${\rm Rm} = \tau_{\rm c} \, \langle {\bm u}^2 \rangle/\eta$ is the magnetic Reynolds number,
$\eta$ is the magnetic diffusion caused by an electrical
conductivity of the plasma, and $\tau_{\rm c}$ is the turbulent correlation time.
Early theoretical predictions by \cite{NS88} 
based on the cumulant expansion method
demonstrated the opposite effect, where the turbulent magnetic diffusion coefficient increases with kinetic helicity, in contradiction to the subsequent numerical results of \cite{BSR17}.

By means of the Feynman diagram technique,
it has been found that kinetic helicity increases the turbulent diffusion of a passive scalar \citep{DS87}.
Later, the increase of passive scalar diffusion of up to 50\% 
by kinetic helicity has been confirmed
by \cite{Chkhetiani+06} when applying the renormalization group approach.
On the other hand, applying the renormalization group theory it has been demonstrated that there is no effect of helicity on the effective eddy viscosity \citep{Zhou90}.
Various helicity effects on different characteristics of turbulence are discussed in the recent review by \cite{PY22}.

In the present study, we apply the path-integral approach 
\citep[see, e.g.,][]{DM84,KRS02,EKRS00,EKRS01}
for a random helical velocity field with a finite correlation time for large fluid and magnetic Reynolds numbers.
We derive equations for the mean magnetic field and the mean scalar field (e.g., the mean particle number density).
We have shown that the turbulent magnetic diffusion coefficient decreases because of the kinetic helicity. 
On the other hand, the kinetic helicity increases the
turbulent diffusion coefficient of the scalar field.
Both effects are of the order of $(H_{\rm K} \tau_{\rm c})^2/\langle {\bm u}^2 \rangle$.

To derive the mean-field equations for the magnetic and scalar fields, we use an exact solution of 
the governing equations (i.e., the induction equation for the magnetic field and the convection--diffusion equation for the scalar field) in the form of a functional integral for an arbitrary velocity field.
The microscopic diffusion can be described by a Wiener random process, and the functional integral implies an averaging over the Wiener random process.
The used form of the exact solution of the governing equations allows us to separate the averaging over
the Wiener random process and a random velocity field.
The derived mean-field equations for the magnetic and scalar fields are generally integro-differential equations. 
However, when the characteristic scale of variation of 
the mean fields is much larger than the correlation length of a random velocity field, second-order equations (in spatial variables) are recovered for the mean fields.

For the derivation of the mean-field equations,
we consider a random helical velocity field with a small yet finite constant renewal time.
Thus, we apply a model with two random processes: the Wiener random process, which describes
the microscopic diffusion and the random velocity field between the renewals.
This model reproduces important features of some real turbulent flows.
For instance, the interstellar turbulence, which is driven
by supernovae explosions, loses memory in the instants of explosions 
\citep[see, e.g.,][]{ZRS90,LST00}. 
Between the renewals, the velocity field can be random with its intrinsic statistics.
To obtain a statistically stationary random velocity field, we assume that the velocity fields between renewals have the same statistics.

This paper is organized as follows.
In Section~\ref{sec2} we outline the governing equations
and the procedure of the derivation of  the equation
for the mean magnetic field.
In Section~\ref{sec3} we derive the equation for the turbulent magnetic diffusion coefficient.
For comparison with the magnetic case, we derive the mean-field equation for the particle number density in Section~\ref{sec4} and obtain an expression for the turbulent diffusion coefficient.
In Section~\ref{sec5} we compare the theoretical predictions with the results of the DNS.
Finally, we draw conclusions in Section~\ref{sec6}.
\\

\section{Governing equations}
\label{sec2}

The magnetic field $ {\bm B}(t,{\bm r}) $ is determined
by the induction equation
\begin{eqnarray}
{\partial {\bm B} \over \partial t}  + ({\bm u} \cdot \bec{\bm \nabla})
{\bm B} = ({\bm B} \cdot \bec{\bm \nabla}) {\bm u} + \eta \Delta {\bm B} ,
\label{T1}
\end{eqnarray}
where ${\bm u}$ is a random velocity field.
For simplicity, we consider an incompressible velocity field.
Below, we derive the equation for the
mean magnetic field in a random helical
velocity field with a finite correlation time
for large fluid and magnetic Reynolds numbers.

Following a previously developed method \citep{DM84,KRS02},
we use an exact solution of Equation~(\ref{T1}) with an initial
condition $ {\bm B}(t=s,{\bm x}) = {\bm B}(s,{\bm x}) $ in the form
of the Feynman--Kac formula:
\begin{eqnarray}
B_{i}(t,{\bm x})  = \Big\langle G_{ij}(t,s,\bec{\xi}(t,s)) \,
B_{j}(s,\bec{\xi}(t,s))\Big\rangle_{\bec{\xi}},
\label{T5}
\end{eqnarray}
where the function $G_{ij}(t,s,\bec{\xi})$ is determined by
\begin{eqnarray}
{d G_{ij}(t,s,\bec{\xi}) \over ds} = N_{ik} \, G_{kj}(t,s,\bec{\xi}) ,
\label{AA6}
\end{eqnarray}
$N_{ij} = \nabla_j u_{i}$ is the velocity gradient matrix, $\, \bec{\tilde\xi} = \bec{\xi} - {\bm x}$, 
and $\langle ... \rangle_{\bec{\xi}}$ denotes averaging over the Wiener paths
\begin{eqnarray}
\bec{\xi}(t,s) = {\bm x} - \int_{0}^{t-s}
{\bm u}[t-\mu,\bec{\xi}(t,\mu)] \,d\mu + \sqrt{2\eta} \, 
{\bec{\rm w}}(t-s) .
\nonumber\\
\label{AA5}
\end{eqnarray}
Here ${\bec{\rm w}}(t)$ is a Wiener random process defined by the properties
$\langle {\bec{\rm w}}(t) \rangle_{_{\bec{\rm w}}}=0$, and $\langle {\rm w}_i(t+\tau) {\rm w}_j(t) \rangle_{_{\bec{\rm w}}}= \tau \delta _{ij} $,
and $\langle \dots \rangle_{_{\bec{\rm w}}} $ denotes averaging over the statistics of the Wiener process.
We use the Fourier transform defined as
\begin{eqnarray}
{\bm B}(t, \bec{\xi}) = \int \exp(i \bec{\xi} \cdot {\bm q})
{\bm B}(s, {\bm q}) \,d{\bm q} .
\label{CC8}
\end{eqnarray}
Substituting Equation~(\ref{CC8}) into Equation~(\ref{T5}), we obtain
\begin{eqnarray}
B_{i}(s, {\bm x}) &=& \int \Big\langle G_{ij}(t,s,\bec{\xi}(t,s))
\, \exp[i \bec{\tilde\xi} \cdot {\bm
q}] \, B_{j}(s, {\bm q}) \Big\rangle_{\bec{\xi}}
\nonumber\\
& & \times \exp(i {\bm q} \cdot {\bm x}) \,d{\bm q} \; .
\label{C8}
\end{eqnarray}
In Equation~(\ref{C8}) we expand the function
$\exp[i \bec{\tilde\xi} \cdot {\bm q}] $ in a Taylor series at $ {\bm q} = 0 ,$ i.e.,
$\exp[i \bec{\tilde\xi} \cdot {\bm q}] = \sum_{k=0}^{\infty}
(1/k!) (i \bec{\tilde\xi} \cdot {\bm q})^{k}$.
Using the identity $ (i {\bm q})^{k} \exp[i {\bm x} \cdot {\bm q}] =\bec{\nabla}^{k} \exp[i {\bm x} \cdot {\bm q}]$
and Equation~(\ref{C8}), we arrive at the expression
\begin{eqnarray}
B_{i}(t, {\bm x}) &=& \biggl\langle G_{ij}(t,s,\bec{\xi})
\biggl[\sum_{k=0}^{\infty} {(\bec{\tilde\xi} \cdot
\bec{\nabla})^{k}  \over  k!}\biggr] \biggr\rangle_{\bec{\xi}}
\nonumber\\
& & \times \int B_{j}(s, {\bm q}) \exp(i {\bm q} \cdot {\bm x})\,d{\bm q} .
\label{BC8}
\end{eqnarray}

The inverse Fourier transform implies that
$B_{j}(s, {\bm x}) = \int B_{j}(s, {\bm q}) \exp(i {\bm q} \cdot {\bm x}) \,d{\bm q}$, so that Equation~(\ref{BC8}) can be rewritten as
\begin{eqnarray}
B_{i}(t, {\bm x}) =  \Big\langle G_{ij}(t,\bec{\xi}) \,
\exp(\bec{\tilde\xi} \cdot \bec{\nabla}) \Big\rangle_{\bec{\xi}} B_{j}(s, {\bm x})  .
\label{A5}
\end{eqnarray}
Equation~(\ref{CC8}) can be formally regarded as 
an inverse Fourier transform of the function $B_{i}(t,\bec{\xi})$.
However, $ \bec{\xi} $ is the Wiener path, which is not a standard spatial variable.
On the other hand, Equation~(\ref{A5}) was also derived in Appendix~A of \cite{KRS02} applying
a more rigorous method; see also \cite{DM84}.
In this derivation, the Cameron--Martin--Girsanov theorem was used.

\section{Mean-field equations for the magnetic field}
\label{sec3}

In this section, we derive the mean-field equation for a magnetic field
using a random helical velocity field with a small yet finite constant renewal time.
These results can also be generalized for a random renewal time 
\citep[see, e.g.,][]{LST00,KRS02,EKRS01}.
Assume that in the intervals $ \ldots (- \tau, 0]; (0, \tau]; (\tau, 2 \tau]; \ldots $
the velocity fields are statistically independent and have the same statistics. This implies that the velocity
field loses memory at the prescribed instants $ t = k \tau$, where $ k = 0, \pm 1, \pm 2, \ldots .$
This velocity field cannot be considered as a
stationary  (in statistical sense) field for small times $\sim \tau$, however, it
behaves like a stationary field for $t \gg \tau$.

The velocity fields before and after renewal
are assumed to be statistically independent.
We use this assumption to decouple averaging into averaging over two time intervals.
In particular, the function
$G_{ij}(t,\bec{\xi})$ in Equation~(\ref{A5}) is determined by the velocity
field after the renewal, while the magnetic field
$B_{j}(s, {\bm x})$ is determined by the velocity field before renewal.

In Equation~(\ref{A5}) we specify instants $t = (m + 1) \tau$ and $s = m \tau$, and average it over random velocity field, which yield the equation for the mean magnetic field $\meanBB$ as
\begin{eqnarray}
\meanB_i[(m + 1) \tau,{\bm x}] = P_{ij}(\tau,{\bm x}, i \bec{\nabla}) \, \meanB_{j}(m \tau,{\bm x}),
\label{NC11}
\end{eqnarray}
where $\meanB_i[(m + 1) \tau,{\bm x}]=\langle B_{i}((m + 1) \tau, {\bm x})  \rangle_{\bm u}$,
$\meanB_{j}(m \tau,{\bm x})=\langle  B_{j}(m \tau,{\bm x}) \rangle_{\bm u}$,
and
\begin{eqnarray}
P_{ij}(\tau,{\bm x}, i \bec{\nabla}) = \langle\langle G_{ij}(\tau,\bec{\xi})
\, \exp[\bec{\tilde\xi} \cdot \bec{\nabla}] \rangle_{\bec{\xi}} \rangle_{\bm u}  .
\label{NC25}
\end{eqnarray}
Here the time $s=m \tau $ is the last renewal time before $t=(m + 1) \tau$
and $t - s = \tau$. Averaging of the functions
$G_{ij}(\tau,\bec{\xi}) \, \exp[\bec{\tilde\xi}(\tau) \cdot \bec{\nabla}] $
and $B_{j}(m \tau,{\bm x}) $ over random velocity field $\langle ...\rangle_{\bm u}$
can be decoupled into the product of averages
since $B_{j}(m \tau,{\bm x}) $ and
$G_{ij}(\tau,\bec{\xi}) \, \exp[\bec{\tilde\xi}(\tau) \cdot \bec{\nabla}] $ are statistically
independent. Indeed, the field $B_{j}(m \tau,{\bm x})$ is determined in
the time interval $ (- \infty, m \tau] ,$ whereas the function
$G_{ij}(\tau,\bec{\xi}) \, \exp[\bec{\tilde\xi}(\tau) \cdot \bec{\nabla}] $ is defined on the
interval $ (m \tau, (m + 1) \tau] .$ Due to a renewal, the velocity
field as well as its functionals $B_{j}(m \tau,{\bm x})$ and
$G_{ij}(\tau,\bec{\xi}) \, \exp[\bec{\tilde\xi}(\tau) \cdot \bec{\nabla}] $ in these two time
intervals are statistically independent \citep[see][for details]{DM84,KRS02}.

Considering a very small renewal time  and expanding into Taylor series the functions $G_{ij}(\mu,\bec{\xi})$ and
$\exp[\bec{\tilde\xi}(\mu) \cdot \bec{\nabla}]$ entering in $P_{ij}(\mu,{\bm x}, i \bec{\nabla})$
(see Equation~(\ref{NC25})), we obtain
\begin{eqnarray}
&& P_{ij}(\mu,{\bm x}, i \bec{\nabla}) = \Big\langle \Big\langle \Big(\delta_{ij} + \mu N_{ij} + {\mu^2 \over 4} N_{ik} N_{kj} + ...\Big)
\nonumber \\
&& \; \times
\Big(1 + \tilde\xi_m \nabla_m + {1 \over 2} \tilde\xi_m \tilde\xi_n \nabla_m\nabla_n + ...\Big) \Big\rangle_{\bec{\xi}} \Big\rangle_{\bm u} .
\label{NC34}
\end{eqnarray}
Here we take into account that the solution of Equation~(\ref{AA6}) can be written as
\begin{eqnarray}
&& G_{ij}(\mu) = \delta_{ij} + \int_{0}^{\mu} N_{ij}(\mu') \, d\mu' 
\nonumber\\
&& \quad + {1 \over 2} \int_{0}^{\mu} N_{ik}(\mu') \, d\mu' \, \int_{0}^{\mu'} N_{kj}(\mu'') \, d\mu'' + ...
\label{NAA6}
\end{eqnarray}

We consider a random incompressible velocity field with a Gaussian statistics.
We also consider a homogeneous turbulence with the large fluid and magnetic Reynolds numbers.
Therefore,  the operator $P_{ij}(\mu,{\bm x}, i \bec{\nabla})$ is given by
\begin{eqnarray}
&& P_{ij}(\mu,{\bm x}, i \bec{\nabla}) =\delta_{ij} + \mu \langle \langle \tilde\xi_m N_{ij} \rangle_{\bec{\xi}} \rangle_{\bm u} \nabla_m
\nonumber \\
&& \;
+ \biggl\{{1 \over 2} \delta_{ij} \langle \langle \tilde\xi_m \tilde\xi_n \rangle_{\bec{\xi}} \rangle_{\bm u} + {\mu^2 \over 8} \Big[\langle \langle \tilde\xi_m N_{ik} \rangle_{\bec{\xi}} \rangle_{\bm u}
\langle \langle \tilde\xi_n N_{kj} \rangle_{\bec{\xi}} \rangle_{\bm u}
\nonumber \\
&& \;
+ \langle \langle \tilde\xi_m N_{kj} \rangle_{\bec{\xi}} \rangle_{\bm u} \,  \langle \langle \tilde\xi_n N_{ik} \rangle_{\bec{\xi}} \rangle_{\bm u} \Big] \biggr\} \nabla_m\nabla_n + ...,
\label{NC35}
\end{eqnarray}
where we keep only nonzero correlation functions. Now we determine the correlation function
$\langle \langle \tilde\xi_m \tilde\xi_n \rangle_{\bec{\xi}} \rangle_{\bm u}$ for small $\mu$ as
\begin{eqnarray}
&& \langle \langle \tilde\xi_m \tilde\xi_n \rangle_{\bec{\xi}} \rangle_{\bm u} = \mu^2 \langle u_m u_n \rangle + {\mu^4 \over 4} \Big[\langle u_s \nabla_p u_n \rangle \, \langle u_p \nabla_s u_m \rangle
\nonumber \\
&&
+ \langle u_s u_p  \rangle \, \langle (\nabla_p u_n) \,  (\nabla_s u_m) \rangle \Big] + 2 \eta \langle {\rm w}_m {\rm w}_n \rangle_{_{\bec{\rm w}}} ,
\label{NC36}
\end{eqnarray}
where we neglected terms $\sim {\rm O}(\mu^5)$, and hereafter we denote $\langle ... \rangle$ as the averaging over statistics of the random velocity field.

To determine the correlation function $\langle u_i \nabla_p u_j \rangle$,
we use a model for the second moment $\langle u_i({\bm k}) u_j(-{\bm k}) \rangle$ of homogeneous incompressible and helical turbulence in Fourier space in the following form:
\begin{eqnarray}
\langle u_i({\bm k}) u_j(-{\bm k}) \rangle &=& {E_u(k) \over 8 \pi k^2} \biggl[\Big(\delta_{ij} - {k_i \, k_j \over k^2}\Big) \langle {\bm u}^2 \rangle
\nonumber \\
&& \; - {{\rm i} \over k^2} \, \varepsilon_{ijp} \, k_p \, \langle{\bm u} \cdot  {\bm \omega}  \rangle
\biggr] ,
\label{NM15}
\end{eqnarray}
where ${\bm \omega} = \bec{\nabla} {\bm \times} \, {\bm u}$ is the vorticity,  $\delta_{ij}$ is the Kronecker fully symmetric unit tensor,
$\varepsilon_{ijp}$  is the Levy-Civita fully antisymmetric unit tensor,
$\langle{\bm u} \cdot  {\bm \omega}  \rangle$ is the kinetic helicity density,
the energy spectrum function is $E_u(k) = (2/3)  \, k_0^{-1} \, (k / k_{0})^{-5/3}$ in the inertial range of turbulence $k_0 \leq k \leq k_{\nu}$, the wavenumber $k_{0} = 1 / \ell_0$, the length $\ell_0$ is the integral scale of turbulence, the wavenumber $k_{\nu}=\ell_{\nu}^{-1}$, the length $\ell_{\nu} = \ell_0 {\rm Re}^{-3/4}$ is the Kolmogorov (viscous) scale.
After integration in the Fourier space, we obtain that the correlation function $\langle u_i u_j \rangle$ in the physical space is $\langle u_i u_j \rangle = \langle {\bm u}^2 \rangle \, \delta_{ij}/3$.
Using Equation~(\ref{NM15}), after integration in the Fourier space, we arrive at the following expression:
\begin{eqnarray}
\langle u_i \nabla_p u_j \rangle = - {1 \over 6} \varepsilon_{ijp} \, \langle{\bm u} \cdot  {\bm \omega}  \rangle .
\label{NM16}
\end{eqnarray}
Using Equations~(\ref{NC36}) and~(\ref{NM16}), we obtain that the correlation function
$\langle \langle \tilde\xi_m \tilde\xi_n \rangle_{\bec{\xi}} \rangle_{\bm u}$ is given by
\begin{eqnarray}
&& \langle \langle \tilde\xi_m \tilde\xi_n \rangle_{\bec{\xi}} \rangle_{\bm u} = \mu \delta_{mn} \biggl\{2 \eta + {\mu \over 3} \biggl[\langle {\bm u}^2 \rangle - {\mu^2 \over 24} \langle{\bm u} \cdot  {\bm \omega}  \rangle^2 \biggr] \biggr\}.
\nonumber\\
\label{NM17}
\end{eqnarray}
Here we have neglected a small contribution ($\sim \mu^4$) caused by the nonhelical part of turbulence. 
In a similar way, we obtain that the correlation function
$\langle \langle \tilde\xi_i N_{jp} \rangle_{\bec{\xi}} \rangle_{\bm u}$ is given by
\begin{eqnarray}
\langle \langle \tilde\xi_i N_{jp} \rangle_{\bec{\xi}} \rangle_{\bm u} = - \mu \langle u_i \nabla_p u_j \rangle =
 {\mu \over 6} \varepsilon_{ijp} \, \langle{\bm u} \cdot  {\bm \omega}  \rangle .
 \label{NM18}
\end{eqnarray}
Since $\partial \meanBB / \partial t = \lim\limits_{\mu \to 0} [\meanB_i((m + 1) \mu,{\bm x}) - \meanB_i(m \mu,{\bm x})]/\mu$, Equations~(\ref{NC35})--(\ref{NC36}) and~(\ref{NM16})--(\ref{NM18}) yield
the mean-field equation:
\begin{eqnarray}
{\partial \meanBB(t,{\bm x}) \over \partial t} = \alpha {\bm \nabla} \times \meanBB + (\eta + \eta_\mathrm{t}) \Delta \meanBB ,
\label{FNC34}
\end{eqnarray}
where the turbulent magnetic diffusion coefficient is given by
\begin{eqnarray}
\eta_\mathrm{t} = {\tau_{\rm c} \over 3} \left[\langle {\bm u}^2 \rangle - {\tau_{\rm c}^2 \over 3} \, \langle{\bm u} \cdot  {\bm \omega} \rangle^2 \right] ,
\label{FNM20}
\end{eqnarray}
and the $\alpha$ effect is $\alpha=-\tau_{\rm c}\langle{\bm u} \cdot  {\bm \omega} \rangle/3$.
In the derivation of Equations~(\ref{FNC34})--(\ref{FNM20}), we take into account that
$\lim\limits_{\mu \to 0} (\mu \langle{\bm u} \cdot  {\bm \omega}  \rangle)= 2\tau_{\rm c} \, \langle{\bm u} \cdot  {\bm \omega}  \rangle$
and $\lim\limits_{\mu \to 0}  (\mu \langle{\bm u}^2  \rangle)= 2\tau_{\rm c}  \, \langle{\bm u}^2  \rangle$.
Here we also use that ${\rm div} \meanBB = 0$ and $\varepsilon_{ijp} \, \varepsilon_{mnp} = \delta_{im} \delta_{jn} - \delta_{in}\delta_{jm}$.

It has been demonstrated by DNS \citep{BKRY25}, that the correlation time of the turbulent velocity field depends on the kinetic helicity.
It follows from Equation~(\ref{FNM20}) that
\begin{eqnarray}
{\eta_\mathrm{t}(H_{\rm K}) \over  
\eta_\mathrm{t}(0)}
= {\tau_{\rm c}(H_{\rm K}) \over \tau_0} \left[1 - {\tau_{\rm c}^2(H_{\rm K}) \over 3} \, {H_{\rm K}^2 \over \langle {\bm u}^2 \rangle} \right] ,
\label{FNM21}
\end{eqnarray}
where $H_{\rm K}=\langle{\bm u} \cdot  {\bm \omega}\rangle$ is the kinetic helicity density,
$\eta_\mathrm{t}(0) = \eta_\mathrm{t}(H_{\rm K}=0)$ is the turbulent magnetic diffusivity at zero kinetic helicity,
and $\tau_0 =\tau_{\rm c}(H_{\rm K}=0)$ is the correlation time.

\begin{figure}
\centering
\includegraphics[width=8.0cm]{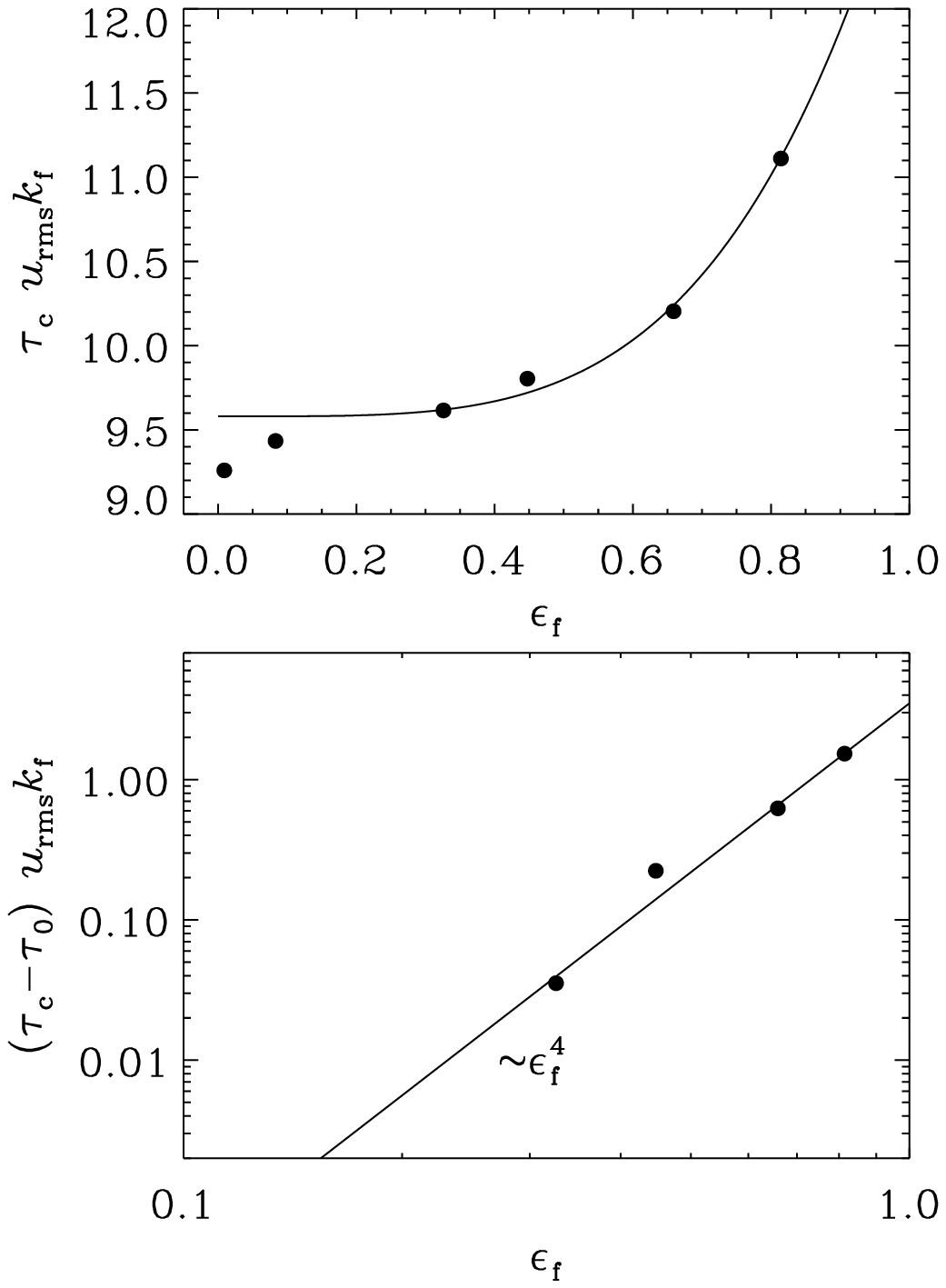}
\caption{\label{Fig2}
Dependence of $\tau_\mathrm{c}$ on $\epsilon_\mathrm{f}$ and $\Rey\approx120$.
The solid line gives the fit with $\zeta=4$ and $C_\tau=0.37$.
In the second panel, we used $\tau_0 u_{\rm rms} k_{\rm f}=9.6$.
}
\end{figure}

We assume that 
\begin{eqnarray}
\tau_{\rm c}(H_{\rm K}) = \tau_0 \, (1 + C_\tau \epsilon_{\rm f}^\zeta) ,
\label{NS21}
\end{eqnarray}
where $\epsilon_{\rm f} = \langle{\bm u} \cdot  {\bm \omega}  \rangle \ell_0/\langle {\bm u}^2 \rangle$ 
is the normalized kinetic helicity.
Equation~(\ref{NS21}) has recently been supported by the DNS of forced turbulence \citep{BKRY25}, where $\zeta=4$ and $C_\tau=0.5$ for $\Rey\approx14$.
This numerical result has been obtained using two independent methods based on the noninstantaneous correlation functions and the rate of energy dissipation.
Equation~(\ref{NS21}) has also been confirmed for $\Rey=120$; see Figure~\ref{Fig2}, where we show 
the dependence of $\tau_\mathrm{c}$ on $\epsilon_\mathrm{f}$.
Here, the simulations had a forcing wavenumber $\kf=5.1\,k_1$, where $k_1$ is the lowest wavenumber in the domain.
In this case, the results are well approximated by $\zeta=4$ and $C_\tau=0.37$, where we have assumed $\ell_0=1/\kf$.

Therefore, the turbulent magnetic diffusion coefficient is
\begin{eqnarray}
\eta_\mathrm{t} =
\eta_\mathrm{t}(0)
\, \left[1 + C_\tau\epsilon_{\rm f}^{4} - \textstyle{1 \over 3} \, \left(1 + C_\tau\epsilon_{\rm f}^{4} \right)^3   \epsilon_{\rm f}^{2} \right] .
\label{FMT2}
\end{eqnarray}
It follows from Equation~(\ref{FMT2}) that the turbulent magnetic diffusion coefficient is reduced by the kinetic helicity (see Section~\ref{sec5}).

\section{Mean-field equation for particle number density}
\label{sec4}

The evolution of the number density  $n(t,{\bm r})$ of small particles advected by
a random incompressible fluid flow is determined by the following convection--diffusion
equation,
\begin{eqnarray}
{\partial n \over \partial t} + {\bm u} \cdot {\bm \nabla} n = \kappa \Delta n ,
\label{PS1}
\end{eqnarray}
where ${\bm u}$ is a random velocity field of the
particles which they acquire in a random fluid velocity
field and $\kappa$ is the coefficient of molecular (Brownian) diffusion.
Following to the method described in Sections~\ref{sec2}--\ref{sec3} \citep[see also][]{EKRS00,EKRS01},
we derive the mean-field equation for the particle number density.
We use an exact solution of Equation~(\ref{PS1}) with an initial
condition $n(t=s,{\bm x}) = n(s,{\bm x}) $ in the form
of the Feynman--Kac formula:
\begin{eqnarray}
n(t,{\bm x})  = \Big\langle n(s,\bec{\xi}(t,s))\Big\rangle_{\bec{\xi}},
\label{PS2}
\end{eqnarray}
where $\langle ... \rangle_{\bec{\xi}}$ implies the averaging
over the Wiener paths:
\begin{eqnarray}
\bec{\xi}(t,s) = {\bm x} - \int_{0}^{t-s}
{\bm u}[t-\mu,\bec{\xi}(t,\mu)] \,d\mu + \sqrt{2\kappa}
\, {\bec{\rm w}}(t-s) .
\nonumber\\
\label{APA5}
\end{eqnarray}
We assume that
\begin{eqnarray}
n(t, \bec{\xi}) = \int \exp(i \bec{\xi} \cdot {\bm q})
n(s, {\bm q}) \,d{\bm q} .
\label{PS3}
\end{eqnarray}
Substituting Equation~(\ref{PS3}) into Equation~(\ref{PS2}), we obtain
\begin{eqnarray}
n(s, {\bm x}) &=& \int \Big\langle \exp[i \bec{\tilde\xi} \cdot {\bm q}] \, n(s, {\bm q}) \Big\rangle_{\bec{\xi}}
\exp(i {\bm q} \cdot {\bm x}) \,d{\bm q}.
\nonumber\\
\label{PS4}
\end{eqnarray}
In Equation~(\ref{PS4}) we expand the function
$\exp[i \bec{\tilde\xi} \cdot {\bm q}] $ in Taylor series at ${\bm q} = 0$ and use the identity $(i {\bm q})^{k} \exp[i {\bm x} \cdot {\bm q}] =\bec{\nabla}^{k} 
\exp[i {\bm x} \cdot {\bm q}]$, which yields
\begin{eqnarray}
n(t, {\bm x}) &=& \biggl\langle \biggl[\sum_{k=0}^{\infty} {(\bec{\tilde\xi} \cdot
\bec{\nabla})^{k} \over  k!}\biggr] \biggr\rangle_{\bec{\xi}} \int n(s, {\bm q}) \exp(i {\bm q} \cdot {\bm x})
\,d{\bm q} .
\nonumber\\
\label{PS6}
\end{eqnarray}
Applying the inverse Fourier transform
$n(s, {\bm x}) = \int n(s, {\bm q}) \exp(i {\bm q} \cdot {\bm x})\,d{\bm q}$, we obtain
\begin{eqnarray}
n(t, {\bm x}) =  \Big\langle \exp(\bec{\tilde\xi} \cdot \bec{\nabla}) \Big\rangle_{\bec{\xi}} n(s, {\bm x})  .
\label{PS5}
\end{eqnarray}
Equation~(\ref{PS5}) has been also derived  applying
a more rigorous method in Appendix~A of \cite{EKRS00}.
In this derivation the Cameron--Martin--Girsanov theorem is applied.

To derive the mean-field equation for a particle number density, we consider a random velocity field with a finite constant renewal time.
In Equation~(\ref{PS5}), we specify instants $t = (m + 1) \tau$ and $s =m \tau$, and average this equation over a random velocity field.
This yields the mean-field equation for the particle number density as
\begin{eqnarray}
\meanN[(m + 1) \tau,{\bm x}] = P(\tau,{\bm x}, i \bec{\nabla}) \, \meanN(m \tau,{\bm x}),
\label{PC11}
\end{eqnarray}
where $\meanN[(m + 1) \tau,{\bm x}]=\langle n((m + 1) \tau, {\bm x})  \rangle_{\bm u}$,
$\meanN(m \tau,{\bm x})=\langle  n(m \tau,{\bm x}) \rangle_{\bm u}$, and
\begin{eqnarray}
P(\tau,{\bm x}, i \bec{\nabla}) = \langle\langle \exp[\bec{\tilde\xi} \cdot \bec{\nabla}] \rangle_{\bec{\xi}} \rangle_{\bm u}  .
\label{PC25}
\end{eqnarray}

We consider a random velocity field with a Gaussian statistics and with
large fluid Reynolds numbers and large P\'eclet numbers.
For a small renewal time, expanding the function
$\exp[\bec{\tilde\xi}(\tau) \cdot \bec{\nabla}]$ into Taylor series, we obtain 
  \begin{eqnarray}
&& P(\mu,{\bm x}, i \bec{\nabla}) = 1 + {1 \over 2} \langle \langle \tilde\xi_m \tilde\xi_n \rangle_{\bec{\xi}} \rangle_{\bm u} \nabla_m\nabla_n + ...,
\label{PS35}
\end{eqnarray}
where 
\begin{eqnarray}
&&\langle \langle \tilde\xi_m \tilde\xi_n \rangle_{\bec{\xi}} \rangle_{\bm u} = \mu \delta_{mn} \biggl\{2 \kappa +   
 {\mu \over 3} \biggl[\langle {\bm u}^2 \rangle - {\mu^2 \over 24} \langle{\bm u} \cdot  {\bm \omega}  \rangle^2
 \biggr] \biggr\}.
  \nonumber \\
\label{PNM17}
\end{eqnarray}
Since $\partial \meanN / \partial t = \lim\limits_{\mu \to 0} [\meanN((m + 1) \mu,{\bm x}) - \meanN(m \mu,{\bm x})]/\mu$,
Equations~(\ref{PC11}), (\ref{PS35}) and (\ref{PNM17}) yield the mean-field equation for the particle number density $\meanN(t,{\bm x})$ as
\begin{eqnarray}
{\partial \meanN(t,{\bm x}) \over \partial t} = (\kappa + \kappa_\mathrm{t}) \Delta \meanN ,
\label{PS34}
\end{eqnarray}
where the turbulent diffusion coefficient is given by
\begin{eqnarray}
\kappa_\mathrm{t} = {\tau_{\rm c} \over 3} \left[\langle {\bm u}^2 \rangle - {\tau_{\rm c}^2 \over 6} 
\, \langle{\bm u} \cdot  {\bm \omega} \rangle^2 \right] .
\label{PNM20}
\end{eqnarray}
It follows from Equation~(\ref{PNM20}) that
\begin{eqnarray}
{\kappa_\mathrm{t}(H_{\rm u}) \over  
\kappa_\mathrm{t}(0)}
= {\tau_{\rm c}(H_{\rm K}) \over \tau_0} \left[1 - {\tau_{\rm c}^2(H_{\rm K}) \over 6} \, {H_{\rm K}^2 \over \langle {\bm u}^2 \rangle} \right] ,
\label{PNM21}
\end{eqnarray}
where $\kappa_\mathrm{t}(0)
= \kappa_\mathrm{t}(H_{\rm K}=0)$ and $\tau_0 =\tau_{\rm c}(H_{\rm K}=0)$.
Since $\tau_{\rm c}(H_{\rm K}) / \tau_0 = 1 + C_\tau \epsilon_{\rm f}^4$ [see Equation~(\ref{NS21})],
the turbulent diffusion coefficient is
\begin{eqnarray}
\kappa_\mathrm{t} =
\kappa_\mathrm{t}(0)
\, \left[1 + C_\tau\epsilon_{\rm f}^{4} - \textstyle{1 \over 6} \, \left(1 + C_\tau\epsilon_{\rm f}^{4} \right)^3  \epsilon_{\rm f}^{2} \right] .
\label{FST3}
\end{eqnarray}
Using Equation~(\ref{FST3}) we will show in Section~\ref{sec5} that the turbulent diffusion coefficient for the scalar field is enhanced by the kinetic helicity.

\begin{figure}
\centering
\includegraphics[width=8.0cm]{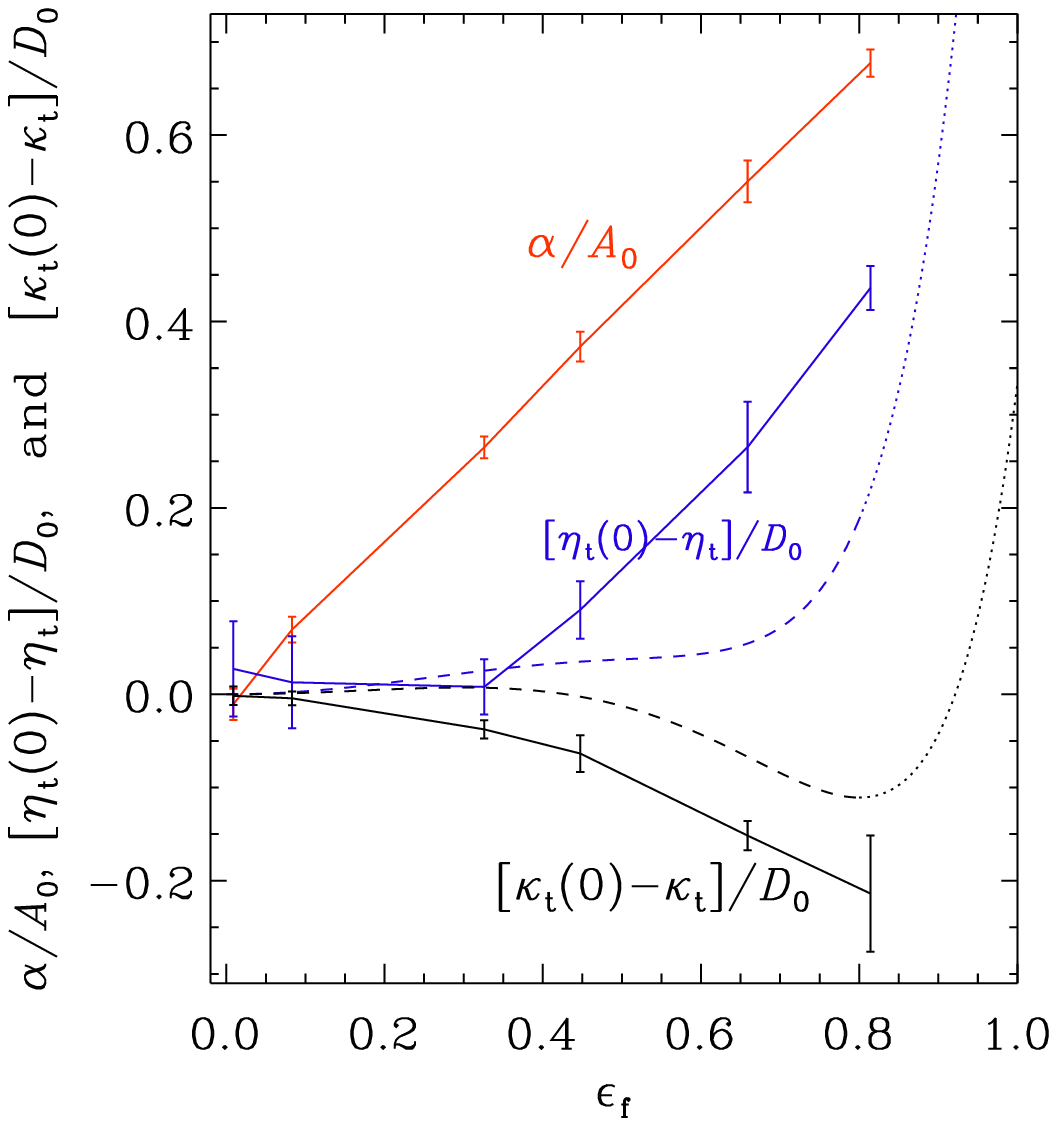}
\caption{\label{Fig1}
Dependencies of $\alpha$ (red solid line), 
$\eta_\mathrm{t}(0) -\eta_\mathrm{t}$ (blue solid line) and
$\kappa_\mathrm{t}(0) -\kappa_\mathrm{t}$ (black solid line)
on the fraction $\epsilon_{\rm f}$ of the kinetic helicity for $\Rey\approx120$.
The theoretical dependencies for $0 <\epsilon_\mathrm{f}< 0.8$ [see Equations~(\ref{FMT2}) and (\ref{FST3})] are shown as dashed lines.
The theoretical results for $\epsilon_\mathrm{f}\ga0.8$ are shown as dotted lines, because they may not be reliable.
}
\end{figure}

\section{Comparisons with numerical results}
\label{sec5}

As in \cite{BKRY25}, we compute a turbulent velocity field by solving the fully compressible momentum equation with an isothermal equation of state.
In \cite{BSR17}, only fully helical cases were compared with nonhelical ones.
\cite{BKRY25} did consider runs with intermediate helicity of the forcing, but only for $\Rey\approx14$.
Here, we also compute such cases with $\Rey\approx120$.
This is accomplished by adding a fraction $\sigma ik_p\epsilon_{ijp} f_j$ to the nonhelical forcing function $f_i$.

\begin{table*}\caption{
Values of $\kappat$ and $\etat$ normalized by $D_0\equiv\urms/3\kf$,
as well as $\alpha$ normalized by $A_0\equiv\urms/3$
for $\Rey\approx14$ and $\approx120$ for different values of $\sigma$.
The values of $\Ma$, $\omega_\mathrm{rms}/\urms\kf$ and $\tau_\mathrm{c}\urms\kf$ are also given.
}\hspace{-10mm}\vspace{12pt}\centerline{\begin{tabular}{lccccccccccccc}
Run & $\Rey$ & $\sigma$ & $2\sigma/(1+\sigma^2)$ & $\epsilon_\mathrm{f}$ & $\kappat/D_0$ & $\etat/D_0$ & $\alpha/A_0$ &
$\Ma$ & $\omega_\mathrm{rms}/\urms\kf$ & $\tau_\mathrm{c}\urms\kf$ \\
\hline
A & $ 13.8$ & $ 0.10$ & $ 0.20$ & $ 0.20$ & $   2.40\pm  0.00$ & $   2.00\pm  0.00$ & $  -0.99\pm  0.01$ & $0.099$ & $  6.1$ & $0.107$ \\
B & $ 14.0$ & $ 0.20$ & $ 0.38$ & $ 0.38$ & $   2.42\pm  0.00$ & $   1.94\pm  0.01$ & $  -1.92\pm  0.00$ & $0.100$ & $  6.0$ & $0.104$ \\
C & $ 14.2$ & $ 0.30$ & $ 0.55$ & $ 0.54$ & $   2.48\pm  0.00$ & $   1.86\pm  0.00$ & $  -2.74\pm  0.01$ & $0.101$ & $  5.9$ & $0.099$ \\
D & $ 14.5$ & $ 0.40$ & $ 0.69$ & $ 0.67$ & $   2.53\pm  0.00$ & $   1.73\pm  0.01$ & $  -3.39\pm  0.00$ & $0.103$ & $  5.8$ & $0.093$ \\
E & $ 14.8$ & $ 0.50$ & $ 0.80$ & $ 0.76$ & $   2.59\pm  0.00$ & $   1.61\pm  0.01$ & $  -3.82\pm  0.02$ & $0.106$ & $  5.7$ & $0.087$ \\
F & $ 15.4$ & $ 0.70$ & $ 0.94$ & $ 0.87$ & $   2.81\pm  0.01$ & $   1.44\pm  0.00$ & $  -4.28\pm  0.04$ & $0.110$ & $  5.5$ & $0.076$ \\
G & $ 15.8$ & $ 1.00$ & $ 1.00$ & $ 0.91$ & $   2.88\pm  0.01$ & $   1.37\pm  0.02$ & $  -4.44\pm  0.05$ & $0.113$ & $  5.3$ & $0.071$ \\
\hline
H & $120.6$ & $ 0.00$ & $ 0.00$ & $ 0.01$ & $   2.27\pm  0.01$ & $   1.73\pm  0.05$ & $   0.05\pm  0.09$ & $0.123$ & $ 13.0$ & $0.054$ \\
I & $121.1$ & $ 0.05$ & $ 0.10$ & $ 0.08$ & $   2.27\pm  0.01$ & $   1.75\pm  0.05$ & $  -0.35\pm  0.07$ & $0.124$ & $ 12.9$ & $0.053$ \\
J & $121.5$ & $ 0.20$ & $ 0.38$ & $ 0.33$ & $   2.30\pm  0.01$ & $   1.75\pm  0.03$ & $  -1.35\pm  0.06$ & $0.124$ & $ 12.8$ & $0.052$ \\
K & $121.5$ & $ 0.30$ & $ 0.55$ & $ 0.45$ & $   2.33\pm  0.02$ & $   1.67\pm  0.03$ & $  -1.90\pm  0.08$ & $0.124$ & $ 12.7$ & $0.051$ \\
L & $124.0$ & $ 0.50$ & $ 0.80$ & $ 0.66$ & $   2.42\pm  0.02$ & $   1.49\pm  0.05$ & $  -2.81\pm  0.11$ & $0.126$ & $ 12.5$ & $0.049$ \\
M & $127.6$ & $ 1.00$ & $ 1.00$ & $ 0.81$ & $   2.48\pm  0.06$ & $   1.32\pm  0.02$ & $  -3.45\pm  0.07$ & $0.130$ & $ 12.2$ & $0.045$ \\
\label{TSummary_kf}\end{tabular}}\end{table*}

We compute the turbulent transport coefficient $\alpha$, $\eta_\mathrm{t}$, and $\kappa_\mathrm{t}$ in Equations (\ref{FNC34}) and (\ref{PS34}) from the turbulent velocity field discussed above.
We use the test-field method \citep{Sch05,Sch07,Bra05,BRS08}, where we solve numerically the equations for the fluctuating magnetic and passive scalar fields.
These are nonlinear inhomogeneous equations, in which the product of the mean magnetic and passive scalar fields acts as an inhomogeneous source term.
Thus, the test-field equations are different from the original evolution equations, which are homogeneous.
Moreover, the mean magnetic and passive scalar fields are not solutions to these equations,
but consist of a set of mutually orthogonal fields that are called test fields.
They are constructed such that we can compute the desired transport coefficient exactly and not as a fit or by some regression method \citep{BS02, Simard+16, Bendre+24}.

The resulting turbulent transport coefficients depend on time and one or two space coordinates (here only on $z$, in addition to $t$).
We are usually interested in their averaged values.
To determine error bars, we also compute averages for any one-third of the full time series.
The results for $\Rey\approx14$ and $\approx120$ are given in Table~\ref{TSummary_kf} for different values of $\sigma$.
For $\sigma<0.7$, $\epsilon_{\rm f} = \langle{\bm u} \cdot  {\bm \omega}  \rangle \ell_0/\langle {\bm u}^2 \rangle$
is well approximated by $2\sigma/(1+\sigma^2)$.
For larger values, $\epsilon_{\rm f}$ stays somewhat below this estimate.
This departure contributes to the steep power-law scaling with $\zeta=4$.

The values $\alpha$, $\eta_\mathrm{t}$, and $\kappa_\mathrm{t}$ for $\Rey=120$ are plotted in Figure~\ref{Fig1}.
As in \cite{BKRY25}, we present them in normalized form
and divide $\alpha$ by $A_0=u_\mathrm{rms}/3$ and $\eta_\mathrm{t}$ and $\kappa_\mathrm{t}$ by $D_0=u_\mathrm{rms}/3k_\mathrm{f}$.
Note that $\eta_\mathrm{t}(0)=\kappa_\mathrm{t}(0)=D_0$.
We see that $\alpha$ increases approximately linearly with $\epsilon_\mathrm{f}$.
For $\eta_\mathrm{t}$ and $\kappa_\mathrm{t}$, it is convenient to plot the differences from the nonhelical values, $\eta_\mathrm{t}(0)$ and $\kappa_\mathrm{t}(0)$, respectively.
We see that for both functions, the differences are small when $\epsilon_\mathrm{f}\la0.4$, and then depart from zero in opposite directions.
This is also predicted by the theory.
For $\epsilon_\mathrm{f}\ga0.8$, however, there are major departures between our theory and the simulations.
Note that the simulations \citep{BKRY25} predict similar results both for passive scalars using the test-field method and for active scalars based on the decay of an initial entropy perturbation. 

The strong dependence of the theoretical results from Equations~(\ref{FMT2}) and (\ref{FST3}) involving high powers of $\epsilon_\mathrm{f}$ 
is related to the following reasons.
The main contributions to the difference in turbulent diffusion coefficients for helical and nonhelical turbulence come from the fourth-order moments of a random velocity field. 
The second reason for the high powers of $\epsilon_\mathrm{f}$ in turbulent diffusion coefficients is related to the strong dependence
of the correlation time of a random velocity field on $\epsilon_\mathrm{f}$ found in simulations.

The difference between the theoretical predictions and the simulations for $0.8 < \epsilon_\mathrm{f} \leq 1$ is related to the theory being based on the following assumptions:
(i) the contributions of higher than fourth-order moments of a random velocity field are neglected;
(ii) it is assumed that the velocity field has Gaussian statistics; and
(iii) we use a model of a random velocity field with renovations.

A recent theoretical study by \cite{KS25}, where a method based on the Furutsu--Novikov theorem \citep{Furutsu63, Novikov65} has been applied, shows that the turbulent diffusivities of both the mean magnetic and passive scalar fields are suppressed by kinetic helicity.
In that paper, however, the kinetic helicity dependence of the correlation time has not been taken into account 
\citep{KS25}.
This may explain the discrepancy with the numerical results
related to the helicity effect on turbulent diffusion of the scalar field \citep{BKRY25}.

\section{Discussion}
\label{sec6}

One of the main effects of astrophysical turbulent flows is a strong increase in the diffusion
of the large-scale magnetic and scalar fields, which can be characterized in terms of the effective (turbulent)
diffusion coefficients. 
The latter effect decreases the growth rates of the mean-field dynamo instability
and various clustering instabilities related to scalar fields.
 
In the present study, we have developed a theory that
qualitatively
explains the nontrivial behavior of turbulent diffusion coefficients of the large-scale magnetic and scalar fields as functions of the kinetic helicity.
These effects have been recently discovered by DNS \citep{BSR17,BKRY25}, which show that turbulent magnetic diffusion decreases
with increasing kinetic helicity while turbulent diffusion of passive scalars increases with the helicity.

The main contribution to these effects comes from the fourth-order correlation function of the turbulent velocity field.
This is the reason why widely used methods like 
the quasi-linear approach [the first-order smoothing approximation (FOSA) or the second-order correlation approximation (SOCA)] as well as the various $\tau$ 
approaches and the direct interaction approximation (DIA) cannot describe these effects.
For instance, \cite{GD95} use the quasi-linear approach to determine the turbulent transport coefficients
(the $\alpha$ effect and turbulent magnetic diffusivity).
The main assumption of the quasi-linear approach
is that fluctuations are much smaller than the mean fields,
so the fourth-order moments have been neglected by
\cite{GD95}.
All studies of the kinetic helicity effect on turbulent diffusivity discussed here apply various perturbation approaches that take into account the
fourth-order moments of random Gaussian velocity fields with small yet finite correlation times.

The main goal of the present paper is to explain the results of the numerical simulations by \cite{BSR17} and \cite{BKRY25}, where we also take into account
the kinetic helicity dependence of the correlation time of the random velocity field that has been found in DNS.
We have applied the path-integral approach
for random flows with a finite correlation time and for large Reynolds and P{\'e}clet numbers. 
We have assumed that the velocity field has Gaussian statistics,
which allows us to represent the fourth-order moments of the turbulent velocity field as a product of second-order moments.
A crucial role in the understanding of these effects is played by the kinetic helicity effect 
on the turbulent correlation time, which increases with increasing helicity.
The results of the theory developed here are in a qualitative
agreement with the numerical results of \cite{BSR17} and \cite{BKRY25}.

\begin{acknowledgements}
We thank Petri J.\ K\"apyl\"a and Nobumitsu Yokoi
for very useful discussions.
We also acknowledge the discussions with participants
of the Nordita Scientific Program on ''Stellar Convection: Modelling, Theory and Observations", Stockholm (September 2024).
This research was supported in part by the
Swedish Research Council (Vetenskapsr{\aa}det) under Grant No.\ 2019-04234,
the National Science Foundation under grant no.\ NSF AST-2307698, a NASA ATP Award 80NSSC22K0825.
We acknowledge the allocation of computing resources provided by the
Swedish National Allocations Committee at the Center for
Parallel Computers at the Royal Institute of Technology in Stockholm.

\vspace{2mm}\noindent
{\em Software and Data Availability.} The source code used for
the simulations of this study, the {\sc Pencil Code} \citep{JOSS},
is available on \url{https://github.com/pencil-code/}.
The simulation setups and corresponding secondary data are available on
\dataset[http://doi.org/10.5281/zenodo.15084461]{http://doi.org/10.5281/zenodo.15084461}.
\end{acknowledgements}

\bibliography{ref}{}
\bibliographystyle{aasjournal}
\end{document}